\documentclass[aip,jcp, graphicx, reprint]{revtex4-1}

\usepackage{chemformula} 
\usepackage[T1]{fontenc} 
\usepackage{multirow}
\usepackage{comment}
\usepackage{pdfpages}

\draft 

\makeatletter
\AtBeginDocument{\let\LS@rot\@undefined}
\makeatother

\begin{document}
%

%
\begin{table}
  \begin{tabular}{|c|}
    \hline
~~~{\normalsize This article may be downloaded for personal use only. 
Any other use requires prior permission of the author}~~~\\ 
{\normalsize and AIP Publishing. This article appeared in
\textit{The Journal of Chemical Physics}, 157, 024504 (2022)}\\
{\normalsize and may be found at:~~~}\texttt{\normalsize https://doi.org/10.1063/5.0093958}\\
 \hline
  \end{tabular}
  \end{table}

\vspace*{0cm}   
~\\~\\

\title{Accurate diffusion coefficients of the excess proton and hydroxide in water         
via extensive \textit{ab initio} simulations with different schemes}



\author{Daniel Mu\~noz-Santiburcio}
\email{daniel.munozsan@upm.es}
\affiliation{CIC nanoGUNE BRTA, Tolosa Hiribidea 76, 20018 San Sebasti\'an, Spain}
\affiliation{Instituto de Fusi\'on Nuclear ``Guillermo Velarde'',
Universidad Polit\'ecnica de Madrid,
C/ Jos\'e Guti\'errez Abascal 2, 28006 Madrid, Spain}


\date{\today}

\begin{abstract}
Despite its simple molecular formula,
obtaining an accurate \textit{in silico} description of water is far from straightforward.
Many of its very peculiar properties are quite elusive,
and in particular obtaining 
good
estimations of the diffusion coefficients
of the solvated proton and hydroxide at a reasonable computational cost
has been an unsolved challenge until now.
Here I present extensive results
of several unusually long \textit{ab initio} MD simulations employing
different combinations of the Born--Oppenheimer
and 2$^\mathrm{nd}$-generation Car--Parrinello MD propagation methods
with different ensembles (NVE and NVT) and thermostats,
which show
that these methods together with the
RPBE-D3 functional provide
a very accurate estimation of the diffusion coefficients of the
solvated H$_3$O$^+$ and OH$^-$ ions,
together with an extremely accurate description
of several properties of neutral water
(such as the structure of the liquid and its
diffusion and shear viscosity coefficients).
In addition, I show that the estimations of $D_\mathrm{H_3O^+}$ and $D_\mathrm{OH^-}$
depend dramatically on the simulation length, being necessary to reach timescales
in the order of hundreds of picoseconds to obtain reliable results.
\end{abstract}


\maketitle 



%
%

%

\section{Introduction}
\label{secintro}

One of the most important and intriguing phenomena occurring in water
is the diffusion of its two autodissociation products, namely the solvated proton and hydroxide ions.
As it is reflected in classical textbooks on electrolytes,\cite{robinson}
the first measurements of their abnormally high diffusion coefficients were seen with fascination,
and it was clear since then that the explanation of this fact would be 
intimately linked to the nature of water itself.\cite{bernal_jcp1933}
While it is now accepted that the diffusion of these 
species occurs via the Grotthuss mechanism,\cite{marx_cpc2006}
which involves consecutive proton transfers between neighboring molecules,
even after many decades of intense investigation 
the particular details of such structural migration mechanism and the nature of the solvated
proton and hydroxide are still a matter of debate.\cite{thamer_science2015, yu_jacs2017, 
	dahms_science2017, chen_natchem2018, fournier_natchem2018, 
	napoli_jcp2018, yuan_acscentsci2019, calio_jacs2021,arntsen_jcp2021}

As seen in those studies, interpreting the experimental results is not straightforward,
and it is necessary to employ computer simulations to obtain a complete picture of the
diffusion process of these species.
However, 
despite the vast computational work already done,\cite{marx_cpc2006,marx_chemrev2010,agmon_chemrev2016,sakti_wires2019}
an  accurate and affordable method capable of reproducing the experimental diffusion coefficients of the  H$^+$(aq) and OH$^-$(aq) was yet to be found.
Indeed, water is one of the most difficult substances to accurately model \textit{in silico},
and even finding a reliable approach for simulating pure water is being a challenge in itself.
In particular, within the \textit{ab initio} molecular dynamics (AIMD) community 
there is a long-standing and ongoing quest for an ideal exchange-correlation (XC) 
functional that is capable of reproducing all of its most important properties.\cite{gillan_jcp2016}
The performance of a given XC functional for liquid water is usually assessed
by evaluating static properties related to the structure of the liquid,
like its radial distribution functions $g_\mathrm{OO}(r)$,  $g_\mathrm{OH}(r)$ and  $g_\mathrm{HH}(r)$,
and the equilibrium density $\rho$ at a given temperature and pressure, 
with dynamical properties such as the diffusion coefficient $D_\mathrm{H_2O}$ in second place.
Ideally, obtaining a proper computational description of the 
H$^+$(aq) and OH$^-$(aq) implies finding an approach that
reproduces their experimental diffusion coefficients \textit{in addition}
to reproducing the  diffusion coefficient of neat water.
Thus, it is worth to start the search of such optimal approach 
for H$^+$(aq) and OH$^-$(aq) by reviewing the 
performance of the different XC functionals for reproducing 
the experimental diffusivity of H$_2$O.
%


For a long time, \textit{ab initio} simulation of water was mostly done employing
bare GGA XC functionals, mainly BLYP~\cite{lee_prb1988, becke_pra1988} and PBE~\cite{Perdew1996+Erratum}
(for a detailed review on these and other functionals for simulations of water, 
including comprehensive tables summarizing results that will be commented in the following,
see Ref.~\citenum{gillan_jcp2016}).
A nagging problem of PBE --and of BLYP to a lesser extent--
known since long ago~\cite{fernandez_jcp2004}
is that it overstructures water, 
and several simulations using indistinctly light or heavy water 
have established that its diffusion coefficients $D$ around 300~K 
are in the order of 
0.2--0.6 $\times 10^{-9}$~m$^2$s$^{-1}$ for BLYP and
0.1--0.3 $\times 10^{-9}$~m$^2$s$^{-1}$ for PBE,\cite{gillan_jcp2016}
roughly one order of magnitude smaller
than the experimental values of both 
H$_2$O and D$_2$O
(respectively $2.39 \times 10^{-9}$~m$^2$s$^{-1}$ 
and 
$1.96 \times 10^{-9}$~m$^2$s$^{-1}$
at 300~K and 1 bar\cite{hardy_jcp2001}).
This issue has been usually remedied by increasing the simulation temperature,
as suggested since the very discovery of the problem.\cite{fernandez_jcp2004}

The development of different dispersion correction schemes
represented a very good way of improving the GGA results at a negligible computational cost. 
While adding dispersion corrections in different forms to the BLYP functional greatly improves 
its description of water and remediates the overstructuring problem~\cite{gillan_jcp2016}
(though the reported $D$ for heavy water is at most about $1.7 \times 10^{-9}$~m$^2$s$^{-1}$ at 308~K
or even higher $T$, thus still somewhat below the experimental value),
the situation is more intricate for PBE.
The DRSLL dispersion correction scheme~\cite{dion_prl2004}
 does soften the structure of PBE water~\cite{wang_jcp2011,corsetti_jcp2013,bankura_jpcc2014}
and improves the $D_\mathrm{D_2O}$ values towards 1.7--2.1$\times 10^{-9}$~m$^2$s$^{-1}$ at 301--304~K,\cite{wang_jcp2011,corsetti_jcp2013}
while TS only improves modestly the radial distribution functions 
without increasing much the low diffusivity ($D_\mathrm{D_2O} = 0.44 \times 10^{-9}$~m$^2$s$^{-1}$ at 300~K,\cite{distasio_jcp2014}) 
and no improvement in either is obtained with 
DCACP~\cite{lilienfeld_prl2004} corrections~\cite{lin_jctc2012}
or with the popular D2~\cite{lin_jctc2012} and D3 corrections.\cite{bankura_jpcc2014} 
Actually, the fact that D2/D3 corrections are contraindicated for PBE simulations of water was already demonstrated 
at the cluster level in the seminal paper 
where
Grimme \textit{et al.}
introduced the D3 correction scheme,\cite{grimme_jcp2010} as seen in Fig.~8 therein
where PBE-D2/D3 are shown to perform much worse than bare PBE for the WATER27 benchmark set.

Despite the PBE functional is still one of the most used for simulations of water,
probably because its robustness for a wide range of systems 
makes it a good choice for simulations of aqueous interfaces and heterogeneous systems,
in recent years its two modified versions revPBE~\cite{zhang_prl1998} 
and RPBE~\cite{hammer_prb1999}
(often including dispersion corrections and/or a fraction of Hartree-Fock exchange)
are becoming new favorites for simulating water.  
Interestingly, revPBE and RPBE are the only GGA functionals that do not produce a severe overstructuring of the liquid.\cite{gillan_jcp2016}
The bare revPBE functional yields $D_\mathrm{D_2O} = 2.1 \times 10^{-9}$~m$^2$s$^{-1}$ at $T=323$~K,\cite{lin_jctc2012} 
which is in very good agreement with the experiment when taking into account the finite-size effects of the simulation.
Other simulations in heavy water show that adding DCACP~\cite{lin_jctc2012} 
or D3~\cite{bankura_jpcc2014} corrections worsens the diffusivity 
by reducing it with respect to both the bare revPBE and experimental values,
while D2~\cite{lin_jctc2012} or DRSLL~\cite{wang_jcp2011} worsen it by its increase.
On the other hand, more recent simulations~\cite{marsalek_jpcl2017} with light water show that revPBE-D3
yields a very good (finite size-corrected) value of $D_\mathrm{H_2O} =2.22 \times 10^{-9}$~m$^2$s$^{-1}$ at 300~K, 
while an even better estimation ($2.29 \times 10^{-9}$~m$^2$s$^{-1}$) and a remarkably exact description
of the structural and spectral features of water are obtained when 
using revPBE0-D3 (which adds a 25\% of Hartree-Fock exchange over revPBE-D3)
\textit{and} including nuclear quantum effects (NQEs) via path integral MD.
About the importance of including NQEs in molecular simulations of water,
it is worth noting that years ago these were thought to slightly weaken the H--bonds (and thus increase the diffusivity)
on the basis of experimental evidence such as isotope effects in light \textit{vs.} heavy water.
This motivated approximations such as slightly elevating the temperature in order to compensate for the missing NQEs
in simulations with classical nuclei.
However, in recent years it has emerged a more complex picture where NQEs
originate competing effects which both weaken and strengthen the H--bonds,\cite{ceriotti_chemrev2016,markland_natrevchem2018}
depending on the precise features of the aqueous system and on the property of interest
whether these cancel out or else one of them dominates.
At that respect, the results obtained with revPBE-D3 and revPBE0-D3 with and without NQEs~\cite{marsalek_jpcl2017}
show that adding NQEs to revPBE-D3 \textit{reduces} the diffusivity (from $2.22\times 10^{-9}$~m$^2$s$^{-1}$ to $1.60\times 10^{-9}$~m$^2$s$^{-1}$), 
but when added to revPBE0-D3 they are largely compensated by the fraction of exact (HF) exchange which has the opposite effect, 
in such a way that bare revPBE-D3 (which includes \textit{neither}) still provides a very good description of water.
This improved understanding of the impact of NQEs discourages the use of 
crude approximations such as elevating the temperature, 
which may be actually counterproductive.\cite{ruizpestana_jpcl2018,li_jctc2022}
On the other hand, the effect of NQEs on simulations differs depending on the underlying method.
For instance, simulations with neural network potentials trained with the SCAN meta-GGA functional~\cite{sun_prl2015}
showed that the $D_\mathrm{H_2O} = 1.1 \times 10^{-9}$~m$^2$s$^{-1}$ with classical nuclei slightly decreased 
to $1.08 \times 10^{-9}$~m$^2$s$^{-1}$ when including NQEs at 300~K (but more noticeably at higher temperatures);
but simulations with the meta-GGA functional B97M-rV~\cite{mardirossian_jcp2015,mardirossian_jpcl2017} 
showed the opposite effect~\cite{ruizpestana_jpcl2018} 
(increase from $D_\mathrm{H_2O} \sim 2.7 \times 10^{-9}$~m$^2$s$^{-1}$ to $\sim 2.9 \times 10^{-9}$~m$^2$s$^{-1}$ at 300~K, 
values inferred from the plot in Fig.~3 of Ref.~\citenum{ruizpestana_jpcl2018}).
I note in passing that meta-GGA functionals do not perform necessarily better than regular GGA functionals for water,
as shown by the previous examples.

Finally, the RPBE functional is nowadays mostly used together with D3 corrections. 
RPBE-D3 performs very well for different aqueous systems, from pure bulk water~\cite{morawietz2013density,forster2014dispersion,imoto2015water,sakong2016structure}
and aqueous solutions\cite{hellstrom2016concentration,imoto2016toward,imoto2018aqueous}
to interfacial~\cite{tonigold2012dispersive} 
and nanoconfined water,\cite{ruiz_jpcl2019,ruiz_pccp2020,munoz_chemrev2021}
including the liquid/vapor equilibrium~\cite{schienbein_jpcb2018,wohlfahrt_jcp2020}
and the liquid at supercritical conditions.\cite{schienbein_jpcb2018,schienbein_pccp2020,schienbein_anie2020,thomsen_jcp2021}
The diffusivity of RPBE-D3 water was first estimated via 
neural network potentials,\cite{morawietz_pnas2016} 
which reported $D_\mathrm{H_2O}$ in the range 3.0--3.2$\times 10^{-9}$~m$^2$s$^{-1}$ 
for systems of different sizes at 300~K 
and $\rho = 1$ g/cm$^3$ after applying the proper finite-size correction 
(see Fig.~S4 in the SI Appendix of Ref.~\citenum{morawietz_pnas2016}).
More recently, it was also estimated directly via AIMD with a quite thorough protocol
involving averaging over several tens of NVE simulations,\cite{schienbein_pccp2020}
yielding $D_\mathrm{H_2O} =2.0 \times 10^{-9}$~m$^2$s$^{-1}$. 
This disagreement between the direct AIMD estimation and the neural network one 
is very likely due to the underlying training set employed for the neural network potentials.

After having observed the notorious difficulties for nailing down the experimental $D$ of neutral water via AIMD, 
it should come as no surprise that obtaining accurate estimations of $D_\mathrm{H_3O^+}$ and $D_\mathrm{OH^-}$ 
has been a challenging task.
For these, on top of the aforementioned problem of finding a good functional, 
there is the problem of the extremely long simulation times required to properly converge the diffusion coefficients 
in comparison to neutral water: obviously, a simulation with $N$ neutral water molecules yields a statistics for $D_\mathrm{H_2O}$ that is $N$
times that of a simulation of the excess proton/hydroxide, where only one of these species per unit cell is usually employed.

Pioneering AIMD simulations~\cite{tuckerman_accchemres2006} with heavy water showed that the 
bare BLYP functional 
reproduces qualitatively the experimental $D_\mathrm{D_3O^+}$/$D_\mathrm{OD^-}$ ratio,
though their absolute values are below the experimental ones by roughly a factor of 2. 
On the other hand, other functionals like HCTH and PW91 produce qualitatively wrong results 
when comparing $D_\mathrm{D_3O^+}$ to $D_\mathrm{OD^-}$.\cite{tuckerman_accchemres2006}
Interestingly, those three functionals yield qualitatively similar values of $D_\mathrm{D_3O^+}$
and also of  $D_\mathrm{D_2O}$, but differ greatly in the estimation of  $D_\mathrm{OD^-}$ which 
varies more than order one of magnitude. 
The underlying reason is that these different functionals
produce fundamentally different solvation structures of the hydroxide ion, which in turn 
results in qualitatively different structural migration mechanisms.\cite{tuckerman_accchemres2006,marx_chemrev2010}
Very recent simulations~\cite{arntsen_jcp2021} using BLYP-D3 and 
two different empirically-corrected versions of it,
analyzing two replica systems in each case,
yielded different values of $D_\mathrm{H_3O^+}$ in a wide range  
($2.9$--$9.8 \times 10^{-9}$~m$^2$s$^{-1}$).
However, there are no data for the OH$^-$(aq) with the same methods.

The best AIMD estimation so far of 
$D$ for the solvated proton and hydroxide
was recently obtained 
using the PBE0-TS functional and employing heavy water at 330~K,\cite{chen_natchem2018}
obtaining $D_\mathrm{D_3O^+} = 8.3 \times 10^{-9}$~m$^2$s$^{-1}$ and 
 $D_\mathrm{OD^-} = 3.7 \times 10^{-9}$~m$^2$s$^{-1}$, 
in fair agreement with the respective experimental values in heavy water at 298~K
 of 6.7--6.9$\times 10^{-9}$~m$^2$s$^{-1}$ and  3.1--3.2$\times 10^{-9}$~m$^2$s$^{-1}$.
The observed underlying migration mechanisms of the hydronium and hydroxide basically agree
with those proposed years back,\cite{marx_chemrev2010}
though pointing out now a prominent role of concerted proton transfer events in the case of H$^+$ migration,
an issue which is however debated.~\cite{arntsen_jcp2021}
Despite such overall good results,
it must be noted that
PBE0-TS somewhat overestimates the experimental $D$ for the hydronium and hydroxide 
while it underestimates that of neutral water even when using a simulation temperature of 330~K\cite{distasio_jcp2014}
($1.46 \times 10^{-9}$~m$^2$s$^{-1}$ in the D$_2$O simulation at 330~K \textit{vs.} the experimental $1.96\times 10^{-9}$~m$^2$s$^{-1}$ of heavy water at 300~K).
Such simulation temperature is used in Refs.~\citenum{distasio_jcp2014}~and~\citenum{chen_natchem2018}
in order to approximately compensate for the missing NQEs,\cite{distasio_jcp2014}
which implies that their computed $D$ values must be compared to the experimental ones at 300~K.
However, it must be stressed that this approximation 
may actually produce the opposite effect compared to the actual NQEs,\cite{marsalek_jpcl2017,li_jctc2022} as said above.

Regarding the importance of NQEs for modeling the diffusion processes of H$_3$O$^+$ and OH$^-$ in water,
it must be realized first that these are governed by the hydrogen-bond fluctuations of the water molecules in
their second and first solvation shells, respectively, and not by the energetic barriers for proton transfer between the H$_3$O$^+$/OH$^-$ 
and their solvating waters.\cite{marx_chemrev2010}
According to this picture, NQEs must affect H$_3$O$^+$/OH$^-$ diffusion 
mostly through the indirect effect they have in the overall dynamics of water.
Indeed, seminal path integral AIMD simulations~\cite{marx_nature1999,tuckerman_nature2002} 
showed that including NQEs did not change these overall diffusion mechanisms\cite{tuckerman_accchemres2006,marx_chemrev2010}
but only changed some peculiar features of both defects.
In the case of the proton, NQEs facilitate the interconversion of the idealized Zundel and Eigen complexes 
(resp. H$_5$O$_2^+$ and (H$_2$O)$_3$H$_3$O$^+$),\cite{marx_nature1999}
which as said is not the limiting step for the structural diffusion.
For the hydroxide, NQEs favor proton transfer within the transient (H$_2$O)$_3$OH$^-$ complex for a wider range of geometries
compared to the classical case where a more stringent geometrical arrangement is required;\cite{tuckerman_nature2002}
this could slightly favor productive proton transfer (i.e. structural diffusion) without qualitatively changing the mechanism.~\cite{marx_chemrev2010}
Looking at the isotope effects experimentally observed for the diffusion coefficients 
(0.82 for D$_2$O/H$_2$O, 0.72 for D$_3$O$^+$/H$_3$O$^+$ and 0.59 for OD$^-$/OH$^-$, all at 301.2~K;\cite{hardy_jcp2001,halle_fartrans1983}
note that isotope effects in dynamical properties are not entirely due to NQEs 
but also partly due to trivial effects because of the greater masses of the substituted atoms that appear also in the classical limit~\cite{ceriotti_chemrev2016}),
these observations may explain why these do not differ much between neutral water and hydronium 
and differ a bit more in the hydroxide case.
Taking all this into account, 
it is reasonable to expect that NQEs would not change much the relative values
$D_\mathrm{H_3O^+}$/$D_\mathrm{OH^-}$,
$D_\mathrm{H_3O^+}$/$D_\mathrm{H_2O}$ and $D_\mathrm{OH^-}$/$D_\mathrm{H_2O}$,
while predicting their impact on the different $D$ values in  absolute terms
seems mere speculation at this point.
All the information laid out above makes it clear
that there is still quite room left for quantitatively improving the 
computational estimations of the diffusion coefficients
of H$_3$O$^+$ and OH$^-$,
hopefully without requiring expensive approaches such as adding HF exchange 
or explicitly including NQEs.
Motivated by the good results in the literature of RPBE-D3 
for many properties of water in very different situations,
and in particular for its diffusion coefficient at ambient conditions,
I performed extensive AIMD simulations to compute the 
diffusion coefficients of the excess proton and hydroxide ions in bulk water at the RPBE-D3 level.
More precisely, the two main goals reached in this paper are:
($i$) providing reliable estimations of $D_\mathrm{H_3O^+}$ and $D_\mathrm{OH^-}$ 
by performing long and accurate Born--Oppenheimer MD simulations,
and 
($ii$) showing that it is possible to obtain equally reliable estimations
of $D_\mathrm{H_3O^+}$ and $D_\mathrm{OH^-}$ at a much smaller computational cost
exploiting the 2$^\mathrm{nd}$-generation Car--Parrinello MD scheme.

\section{Methods}
\label{secmethods}

All the simulations were performed with CP2K.\cite{cp2k_jcp2020}
Unless otherwise stated, the model system for neutral water
consisted in 128 H$_2$O molecules in a cubic cell with
$L=15.663$~\AA~
(resulting in a water density of 996.556 kg/m$^3$,
matching the experimental value at $T=300$~K and $p=1$~bar~\cite{wagner_jpcrefdat2002}), 
in which a proton was added or removed 
to obtain the water systems with a H$_3$O$^+$ or OH$^-$, respectively.
The electronic structure of the system was treated within the spin-restricted
formulation of Kohn-Sham theory
with the RPBE functional~\cite{hammer_prb1999} plus D3 corrections~\cite{grimme_jcp2010}
which included only the two-body terms and had zero damping.
I employed GTH pseudopotentials~\cite{goedecker_prb1996,hartwigsen_prb1998,krack_theorchemacc2005} 
together with a TZV2P basis set,\cite{vandevondele_cpc2005}
using a 500 Ry density cutoff for the expansion of the auxiliary planewave basis set 
and applying NN50 smoothing for both the charge density and its derivatives.
As shown in the literature,\cite{jonchiere_jcp2011}
NN50 smoothing reduces the spurious numerical effects related to the evaluation of the exchange-correlation
energy on a finite grid, 
and tests done in the present study showed that 
it improves the accuracy of the forces in a way equivalent to a cutoff increase of at least 100~Ry.
In addition, tests for determining the accuracy of the TZV2P basis set in comparison with other 
basis sets available for CP2K and with a planewave basis in the near-infinite limit 
(see Sec.~I of the Supporting Information (SI))
showed that TZV2P is probably the best choice in terms of performance \textit{vs.} accuracy for liquid water,
being almost as accurate as QZV3P and better than the popular DZVP-MOLOPT-SR set.\cite{vandevondele_jcp2007}

Two kinds of \textit{ab initio} MD simulations were carried out:
Born--Oppenheimer (BO)~\cite{marx_hutter_2009}
and 2$^\mathrm{nd}$-generation Car--Parrinello (2genCP)~\cite{kuhne_prl2007,kuhne_wires2014} MD.
In the BOMD runs, I set a threshold for the SCF convergence of $2 \times 10^{-7}$
and a timestep of 0.4~fs; the latter allows an excellent accuracy for integrating the equations of 
motion while also facilitates the SCF convergence w.r.t. larger timesteps as e.g.~0.5~fs
(compensating to some extent the cost of performing more MD steps for the same simulation time).
The BOMD simulations were carried out either in the
NVE ensemble or in the NVT ensemble; in the latter case with
a Nos\'e--Hoover chains thermostat~\cite{martyna_jcp1992} with a chain length of 3,
an order of 9 for the Yoshida integrator and a time constant of 100 fs.
On the other hand, the 2genCPMD runs were carried out either
using its original formulation which implies a Langevin-like thermostatting,
or using other thermostat schemes (Nos\'e--Hoover chains or CSVR~\cite{bussi_jcp2007}). 
In all cases, I used a 0.4~fs timestep for consistency with the BOMD settings.
The full details will be unfolded in the Results section.

In all cases, the diffusion coefficients were determined 
as usual via the Einstein's relation 
$ \mathrm{MSD}(t) = 6 D t $ 
where the mean square displacement (MSD)
was computed by following the position 
of the O atom of the molecule of interest 
(in the case of neutral water, averaging the MSD over all the water molecules in the system).
The plots of every $\mathrm{MSD}(t)$ function 
and the time ranges used for computing the corresponding slopes $6 D$
for all the diffusion coefficients in this work
are disclosed in Sec.~III of the SI.

\section{Results and discussion}

\subsection{Determination of finite-size effects and estimation of the viscosity}
\label{secsize}

As known since long ago,~\cite{dunweg_jcp1993, yeh_jpcb2004} diffusion coefficients in liquid systems estimated via MD 
employing periodic boundary conditions are affected by finite-size effects. 
In the present case of a cubic simulation cell,
the `exact' coefficient $D_0$ in an infinite system
and the diffusion coefficient actually obtained in the simulation 
$D_\mathrm{PBC}$ are related by the expression 
$D_0 = D_\mathrm{PBC} + \frac{ k_\mathrm{B} T \xi }{ 6 \pi \eta L }$,
being $k_\mathrm{B}$ the Boltzmann constant, $T$ the temperature,
the numerical factor $\xi \approx 2.837297$,
$\eta$ the shear viscosity of the solvent, and $L$ the length of the simulation cell.

A convenient way of determining the finite-size correction term
$\frac{ k_\mathrm{B} T \xi }{ 6 \pi \eta L }$ consists in 
computing $D_\mathrm{PBC}$ for different system sizes.
The subsequent linear representation of $D_\mathrm{PBC}$ \textit{vs.} $1/L$ 
readily provides $D_0$ as the $y$-intercept and the correction term
as the slope, from which in turn the shear viscosity $\eta$ of the liquid 
can be estimated.
It must be noted that \textit{a priori} there is no guarantee
that this value of the viscosity will be the same as the experimental one,
and therefore using the latter to directly estimate the 
correction term $\frac{ k_\mathrm{B} T \xi }{ 6 \pi \eta L }$ 
may imply further errors.
On the other hand, MD estimations of $\eta$
do not show any significant system-size dependence.\cite{yeh_jpcb2004}

To this aim, three BOMD runs of neutral water systems were carried out
employing respectively 64, 128 and 256 water molecules in cells
of sizes 12.431~\AA, 15.663~\AA, and 19.734~\AA. 
The three were carried out in the NVT ensemble with a Nos\'e--Hoover chains
thermostat with the settings detailed in the Methods section. 
All were equilibrated during 10 ps, followed by production runs 
of 800~ps, 400~ps and 200~ps respectively for the 64, 128 and 256 H$_2$O cases.

\begin{figure}
\includegraphics[width=0.5\textwidth]{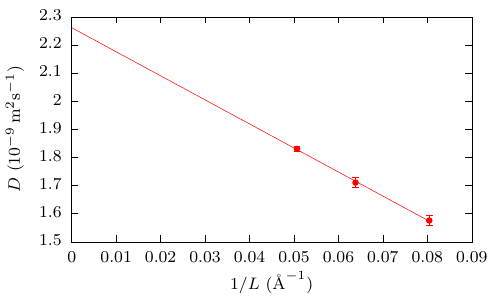}
        \caption{Dependence of the observed diffusion coefficient $D_\mathrm{PBC}$ of neutral water on the
        inverse of the simulation box length $1/L$. The
        error bars indicate the standard deviations of the $D_\mathrm{PBC}$
        computed over the final quarter of the respective simulations.\cite{note_std1}
        The $y$-intercept of the linear fit is $D_0 = 2.26 \times 10^{-9}\,\mathrm{m^2 s^{-1}}$.}
\label{diffvsl}
\end{figure}

As shown in Fig.~\ref{diffvsl}, the dependence of $D_\mathrm{PBC}$
on $1/L$ is linear as expected, providing an estimation of $D_0 = 2.26 \times 10^{-9}\,\mathrm{m^2\,s^{-1}}$
which is in excellent agreement with the experimental $D_\mathrm{exp} = 2.39 \times 10^{-9}\,\mathrm{m^2\,s^{-1}}$
at $T=300$~K.\cite{hardy_jcp2001}
Moreover, the estimation of the shear viscosity that is readily obtained is $\eta_\mathrm{sim} = 7.3 \times 10^{-4}\,\mathrm{Pa \cdot s}$, 
which is in remarkably good agreement with the experimental value $\eta_\mathrm{exp} = 8.9 \times 10^{-4}\,\mathrm{Pa \cdot s}$.~\cite{harris_jced2004}
The resulting additive correction term for the simulations containing 128 water molecules is 
$0.546 \times 10^{-9}$ m$^2$s$^{-1}$
at $T=300$~K.
In the following, all the reported diffusion coefficients 
$D_\mathrm{H_2O}$, $D_\mathrm{H_3O^+}$ and $D_\mathrm{OH^-}$
include this correction term 
$\frac{ k_\mathrm{B} T \xi }{ 6 \pi \eta L }$ 
(where $T$ is the average temperature of each simulation).

\begin{figure}
\includegraphics[width=0.5\textwidth]{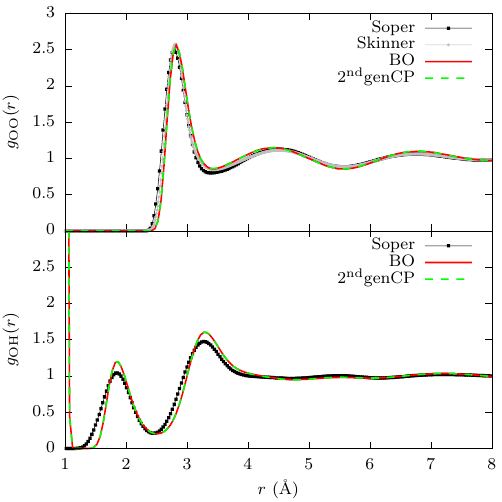}
        \caption{OO and OH radial distribution functions of neutral water 
	obtained in the BO NVT simulation with 128 H$_2$O and in 
	a 2$^\mathrm{nd}$-generation CP simulation (Langevin dynamics with $\gamma_L=\gamma_D =0$),
	in comparison to the experimental results of water at 298~K by Soper~\cite{soper_review2013} 
	and water at 295.1~K by Skinner \textit{et al}.\cite{skinner_jcp2014}}
\label{gofr}
\end{figure}

The previous excellent estimations of the dynamical properties $D$ and $\eta$ of H$_2$O
are accompanied by an accordingly excellent description of static properties such as $g_\mathrm{OO}(r)$ and $g_\mathrm{OH}(r)$,
as shown in Fig.~\ref{gofr}. Finally, I note that the obtained $D_\mathrm{H_2O} = 2.26 \times 10^{-9}$~m$^2$s$^{-1}$ 
is in excellent agreement with that of Ref.~\citenum{schienbein_pccp2020},
which would be $2.1 \times 10^{-9}$~m$^2$s$^{-1}$ if the finite-size correction therein
were computed using $\eta_\mathrm{sim}$ instead of $\eta_\mathrm{exp}$.

\subsection{Diffusion coefficients via BOMD: NVE \textit{vs.} NVT ensembles}
\label{secbo}

In the previous section I have estimated a time-dependent property such as $D$ via 
MD simulations in the NVT ensemble, as it is also usually done in the literature.
However,
it must be kept in mind that the very role of the employed thermostat is `tweaking' the dynamics of the system in order
to maintain the canonical sampling, in such a way that \textit{a priori} there is not any guarantee that the 
resulting dynamics will provide the proper time-dependent properties.
The most accurate way of estimating such observables is to generate 
a set of initial conditions 
(i.e. positions and momenta) representative of a canonical ensemble, 
and with them run many (usually in the order of several tens) microcanonical simulations 
that will indeed provide the proper dynamics of the system,
as done in e.g.~Ref.~\citenum{schienbein_pccp2020} 
(see Fig.~2 therein for a very clarifying scheme of such simulation protocol). 
Averaging the results of that set of simulations will yield the proper
estimation of the desired observable for the canonical ensemble.
However, 
the amount and length of the simulations
required to properly converge this ensemble average
make such procedure cumbersome when done with AIMD,
which is why time-dependent properties are often computed via a single NVT run
or sometimes even a \textit{single} NVE run.

It is frequently assumed that the diffusion coefficients obtained via NVT simulations are acceptable
(except in pathological cases such as when using very strong stochastic thermostats, as I will discuss later),
and indeed in some cases these are remarkably accurate, 
as it is the case of the NVT results for $D_\mathrm{H_2O}$ shown in the previous section.
However, in order to firmly demonstrate that an estimation via NVT is also justified for 
the particular cases of acidic and basic water,
and that the NVT and NVE results are consistent,
I performed 1 NVT and 5 NVE runs for each system 
(including also the neutral water system for the sake of consistency). 
While 5 NVE runs are of course too few to obtain a proper ensemble average, 
they are enough to 
make a semi-quantitative comparison between the NVE and NVT approaches
and discern whether they are in agreement.

In order to obtain the initial conditions for the different NVE runs, 
after an initial equilibration of 20 ps I carried out a NVT run 
with a massive Nos\'e--Hoover chains thermostat (with the same settings described in Sec.~\ref{secmethods}
for the regular (global) Nos\'e--Hoover chains thermostat), 
from which 5 sets of positions and momenta where collected at intervals of 5~ps for each system.
The use of a \textit{massive} Nos\'e--Hoover chains thermostat in this particular case 
guarantees that the sampled sets of positions
and momenta are decoupled from each other even after only 5 ps of time separation.
These initial conditions were used to launch NVE runs of 200~ps length for neutral water and 400~ps for acidic and basic water.
On the other hand, after an initial equilibration of 10 ps I carried out 
a NVT run with a global Nos\'e--Hoover chains thermostat 
of 400 ps length for each system 
(note that for the neutral water case, 
such simulation was already performed during the determination of the finite size effects).

\begin{figure}
        \includegraphics[width=0.5\textwidth]{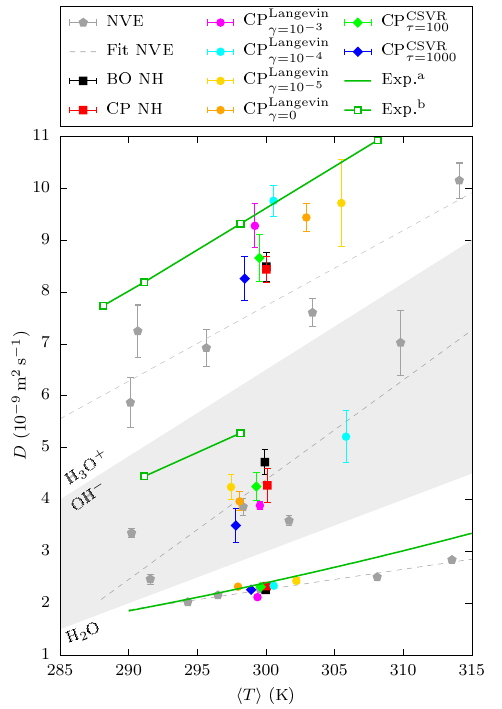}
		\caption{Graphical representation of the diffusion coefficients in Table~\ref{difftab}
		as a function of the average temperature in each simulation. 
		The light/shaded background illustrates the non-overlapping regions where the diffusion coefficients
		of the excess proton, hydroxide and neutral water lie.
		The error bars indicate the standard deviation of $D$, 
		computed over the second half of each simulation.\cite{note_std2}
		Dashed lines are linear fits to the NVE results.
		$\mathrm{^{(a)}\,}$Empirical function for $D_\mathrm{H_2O}$ from Ref.~\citenum{hardy_jcp2001} plotted as a solid line.
		$\mathrm{^{(b)}\,}$Data points for $D_\mathrm{H_3O^+}$ and $D_\mathrm{OH^-}$ from Table~\ref{difftab}, 
		lines between points are a guide to the eye.}
\label{diffplot}
\end{figure}

The obtained diffusion coefficients are listed in Table~\ref{difftab}.
The NVE results are dispersed over a certain range
that is roughly centered around the NVT results,
in such a way that there exists a certain correlation between
the average temperature in each simulation 
and the observed diffusion coefficient $D$,
with a dependence that in the case of  H$_3$O$^+$ and OH$^-$ is quite pronounced
as visualized by the plot of $D$ \textit{vs.}~$\langle T \rangle$ in Fig.~\ref{diffplot}.
This temperature dependence is similar to that shown by the experimental diffusion coefficients
$D_\mathrm{H_3O^+}$ and $D_\mathrm{OH^-}$, 
which is greater than in the case of $D_\mathrm{H_2O}$.
However, despite this general trend,
some NVE simulations with similar $\langle T \rangle$ may yield different $D$ values,
as it occurs for the NVE runs with  $\langle T \rangle \sim 290$~K for both  H$_3$O$^+$ and OH$^-$.
Assuming the best-case scenario that different NVE simulations with the same 
$\langle T \rangle$ must yield the same $D$ in the limit of infinite simulation time,\cite{note_erg}
these results indicate that 400~ps is quite short to properly converge $D_\mathrm{H_3O^+}$ or $D_\mathrm{OH^-}$ 
via microcanonical sampling.
This observation clearly warns against the practice of computing $D$ based on single NVE simulations,
as it is sometimes done in the literature.

Overall,
even though it is evident that several more NVE simulations would be
needed for an accurate determination of the diffusion coefficients via their ensemble-averaging, 
it seems clear that the NVT-predicted coefficients are consistent with the NVE ones.
This conclusion is of course system- and thermostat-dependent, 
and so far it is only completely valid for the precise thermostat settings used in this case.

Regarding the comparison to the experimental results,
the present estimation of $D_\mathrm{H_2O}$ is excellent as already said in Section~\ref{secsize},
while the estimations for the hydronium and hydroxide ions are 
extremely good despite being somewhat below the experimental values. 
However, 
I note that the ratio
$D^\mathrm{exp}_\mathrm{H_3O^+} / D^\mathrm{exp}_\mathrm{OH^-} = 1.8$ at 298~K is
\textit{equal} to the ratio 
$D^\mathrm{sim}_\mathrm{H_3O^+} / D^\mathrm{sim}_\mathrm{OH^-}$ provided by the NVT results at 300~K,
while the ratio $D_\mathrm{H_3O^+} / D_\mathrm{H_2O}$ is  3.9 (experiment) \textit{vs.} 3.8 (simulation),
and the ratio $D_\mathrm{OH^-} / D_\mathrm{H_2O}$ is 2.2 (experiment) \textit{vs.} 2.1 (simulation). 
Therefore,
the current simulation strategy perfectly reproduces the
relative diffusive behavior of all the species
H$_2$O, H$_3$O$^+$ and OH$^-$ in bulk water at ambient conditions.

\begin{table*}
	\caption{Diffusion coefficients (corrected for finite-size effects) and average temperatures for 
	the different simulations of H$_2$O, H$_3$O$^+$(aq) and OH$^-$(aq),
	together with the experimental values at different temperatures.
	All the production simulations had a length of 400~ps, except the H$_2$O simulations in the NVE ensemble which had 200~ps length.
        $D$ is expressed in units of  $10^{-9}$~m$^2$s$^{-1}$ and $T$ in K. 
	$\mathrm{^{(a)}\,}$Data for $D_\mathrm{H_2O}$ obtained from the empirical function in Ref.~\citenum{hardy_jcp2001} (Eq.~3 therein).
	$\mathrm{^{(b)}\,}$Values of $D_\mathrm{H_3O^+}$ and $D_\mathrm{OH^-}$ calculated via Nernst equation $D=\frac{RT\lambda}{F^2}$ 
	using limiting molar conductivities $\lambda$ from Ref.~\citenum{robinson} (Appendix 6.2 therein).}
  \label{difftab}
  \begin{tabular}{|l|l|rrrrrr|}
    \hline
                \multicolumn{2}{|c|}{}      & \multicolumn{6}{c|}{Systems} \\
          \hline
          \multicolumn{2}{|c}{Approach}    & \multicolumn{2}{|c|}{H$_2$O} & \multicolumn{2}{|c|}{ H$_3$O$^+$} & \multicolumn{2}{|c|}{OH$^-$}  \\
          \hline
          Propagation & Ensemble and thermostat  & ~~~~~$D$  & ~~~~$\langle T\rangle$ & ~~~~~$D$  & ~~~~$\langle T\rangle$ & ~~~~~$D$ & ~~~~$\langle T\rangle$  \\
  
  \hline
          \multirow{6}{*}{BO} &  NVE \#1 & 	  2.84 & 313.6&	7.60 & 303.4&	6.64 & 309.8 \\
	  & NVE \#2	&  2.16 & 296.5&	10.15 & 314.1&	2.12 & 291.6\\
	  & NVE \#3 	&  2.28 & 299.6&	5.87&  290.1&	3.22 & 301.7\\
	  & NVE \#4 	&  2.50 & 308.1&	7.25&  290.7&	3.50 & 298.3\\
	  & NVE \#5 	&  2.02 & 294.3&	6.92&  295.7&	3.01 & 290.2\\
	  & NVT Nos\'e--Hoover	&  2.26 & 300.0&	8.49&  300.0&	4.72 & 299.9\\
\hline
	  \multirow{2}{*}{2$^\mathrm{nd}$genCP} &  NVT Langevin $\gamma_L=0.0$ ($\equiv$ NVE) &   2.32 & 298.0&	9.44&  302.9&	3.96 & 298.1\\
	  & NVT Langevin $\gamma_L=10^{-5}$ fs$^{-1}$	&  2.43 & 302.2&	9.71&  305.5&	4.24 & 297.5\\
          & NVT Langevin $\gamma_L=10^{-4}$ fs$^{-1}$ &	  2.34 & 300.6&	9.76&  300.6&	5.21 & 305.8\\
	  & NVT Langevin $\gamma_L=10^{-3}$ fs$^{-1}$ & 	  2.12 & 299.4&	9.28&  299.2&	3.89&  299.6\\
 & NVT Nos\'e--Hoover &	  2.33 & 300.1&	8.43&  300.0&	4.27&  300.1\\
 & NVT CSVR $\tau = 1000$~fs &	  2.26 & 298.9&	8.26&  298.5&	3.50&  297.8\\
 & NVT CSVR $\tau = 100$~fs &	  2.32 & 299.6&	8.65&  299.5&	4.25&  299.3\\
    \hline
	  \multicolumn{2}{|c|}{ \multirow{4}{*}{Experimental$^\mathrm{a,b}$} }    &  2.11  &  295.0     & 7.74   & 288.2   &     &     \\
  \multicolumn{2}{|c|}{}              &  2.29   &  298.2   & 8.19   & 291.2     & 4.45   & 291.2   \\
  \multicolumn{2}{|c|}{}              &  2.39   &  300.0     & 9.31   & 298.2   & 5.28   & 298.2   \\
  \multicolumn{2}{|c|}{}              &  2.69   &  305.0    & 10.93  & 308.2   &     &     \\
\hline
  \end{tabular}
\end{table*}

\subsection{2$^\mathrm{nd}$-generation CP MD: setting the stage}

I have just shown 
that it is possible to 
obtain excellent estimations of the diffusion coefficients
of H$_2$O,  H$_3$O$^+$(aq) and OH$^-$(aq) via Born--Oppenheimer MD in the NVT ensemble, 
though the simulation times involved are rather high considering the high computational cost
of a BOMD simulation.
It is then worthwhile to try alternative approaches 
for performing NVT simulations with Born--Oppenheimer accuracy at a smaller computational cost.
This is precisely the case of the 2$^\mathrm{nd}$-generation Car--Parrinello propagation method,\cite{kuhne_prl2007,kuhne_wires2014}
where the full SCF procedure during each MD step is substituted by a prediction-correction scheme.
At each MD step, this involves the extrapolation of the wavefunction from the previous steps' wavefunctions, 
followed by one or two correction steps that are equivalent to the usual SCF steps in a regular BOMD.
After testing different settings for the stability and performance of the propagation,
I chose to employ an order of 2 for the ASPC extrapolation~\cite{kolafa_jcc2004} and 
a fixed number of 2 corrector steps per MD step;
I note that the convergence of the wavefunction is not enforced in any further way
during the full simulation length. 
This was enough to obtain a $\sim 2.5 \times$ speedup w.r.t. the BOMD simulations.

In its original formulation, the  
2genCP
method allows to perform
the canonical sampling of a system by propagating it with a Langevin-type equation:
\begin{equation}
\label{eq2genCP}
M_I \mathbf{\ddot{R}}_I =  \mathbf{F}_\mathrm{PC} - \gamma_L M_I \mathbf{\dot{R}}_I + \mathbf{\Xi}_I
\end{equation}
being $\mathbf{\Xi}_I$ a random noise term defined by
$\langle \mathbf{\Xi}_I(0) \mathbf{\Xi}_I(t) \rangle = 6 (\gamma_L + \gamma_D) M_I k_\mathrm{B} T \delta (t) $.
The reason behind such approach was that regularly integrating the Newtonian equations of motion 
with the forces obtained via the predictor-corrector scheme $\mathbf{F}_\mathrm{PC}$ led
to a dissipative dynamics (though this effect is system-dependent, being less important in systems like water).\cite{kuhne_prl2007}
Thus, by introducing the friction and white noise terms $- \gamma_L M_I \mathbf{\dot{R}}_I$ 
and $\mathbf{\Xi}_I$ where the latter is defined not by $\gamma_L$ but by $(\gamma_L + \gamma_D)$,
a Langevin dynamics is established where the white noise term is `enhanced' in order to compensate for the intrinsic dissipative dynamics
of the system.
In practice, a convenient way of setting up such simulations is 
choosing a given $\gamma_L$ and then adjusting 
the value of $\gamma_D$ on-the-fly during the equilibration of the system until 
obtaining the desired average temperature.

This method is therefore an excellent alternative to BOMD when it comes to performing
long, expensive MD simulations requiring \textit{ab initio} accuracy, as it is the case of 
the determination of diffusion coefficients. 
However, computing dynamical properties in cases where the underlying dynamics 
is stochastic --as it is the case of Langevin dynamics-- is a delicate issue.
While it has been shown that the 2genCP method provides good estimations of both static and 
dynamical properties of systems like water~\cite{kuhne_jctc2009} 
when using a small $\gamma_L$ ($\sim 10^{-5}$ fs$^{-1}$),
it must be recognized that small values of $\gamma_L$ necessarily imply
longer simulation times until the canonical sampling is completely established,
while on the other hand higher values of $\gamma_L$ eventually destroy the true dynamics of the system,
thus spoiling the estimation of time-dependent properties.

It is therefore desirable to perform a systematic study to find which parameters 
provide the best sampling of the systems of interest.
To that aim, for each of the three studied systems I set up four 2genCP simulations 
using different values of the $\gamma_L$ parameter
(specifically, 0, $10^{-5}$, $10^{-4}$, and $10^{-3}$~fs$^{-1}$)
starting with $\gamma_D = 0$ in all cases and
equilibrating the systems during 10~ps.
Surprisingly, after letting all the simulations to evolve over up to 400~ps of production time,
it was observed that the dynamics of all the studied systems was not dissipative at all,
even though $\gamma_D$ was in all cases kept as 0 during the whole simulation.
Instead, inspection of the running average of the temperature
as a function of the simulation time 
(see plots in Sec.~II of the SI)
showed that the temperature in the
simulations with  $\gamma_L = 0$ tended to stabilize around some value 
not far from 300~K
(similarly to what happened in the BO-NVE simulations),
while the simulations with non-zero  $\gamma_L$ values
behaved as regular NVT runs with different thermostat strengths.
On the other hand, the fluctuations of the instantaneous temperature w.r.t.
the average value (quantified as the variance $\sigma^2_T$, see Sec.~II in the SI) 
in the  $\gamma_L = 0$ cases were similar
to those observed in the BO-NVE simulations, 
while in the  $\gamma_L > 0$ cases they approached 
the ideal value of $\sigma^2_T$ for a proper NVT simulation~\cite{frenkel_smit}
as  $\gamma_L$ increased.
This behavior was consistently observed for all the systems --neutral, acidic and basic water--.
In addition, static properties of water like $g_\mathrm{OO}(r)$ and $g_\mathrm{OH}(r)$ 
were identical to those obtained with BOMD
(Fig.~\ref{gofr}),
while the $D$ values were comparable to the BOMD ones
for H$_2$O, H$_3$O$^+$ and OH$^-$,
with only small differences that are easily explained by the different values of 
$\gamma_L$ as will be discussed in the next section.

\begin{figure}
        \includegraphics[width=0.5\textwidth]{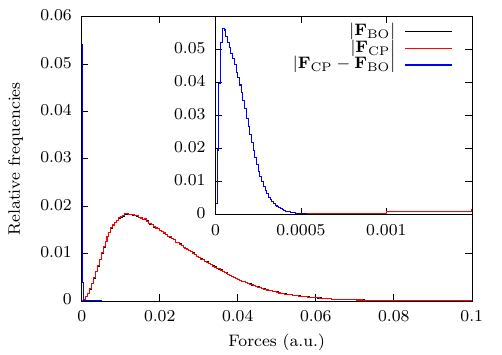}
        \caption{Histogram of the force modules obtained for 4000 snapshots sampled during the 
	2genCP run
	of neutral water with Langevin dynamics with $\gamma_L=\gamma_D=0$ (red), 
	in comparison with the corresponding ones obtained at the Born--Oppenheimer surface 
	(black, practically coincident with the red histogram)
	and with the module of the difference between the 2genCP and BO forces (blue).
	The inset shows a magnification of the latter. 
	The histograms obtained for the acidic and basic water systems are indistinguishable from these.} 
\label{histo}
\end{figure}

\begin{figure}
        \includegraphics[width=0.5\textwidth]{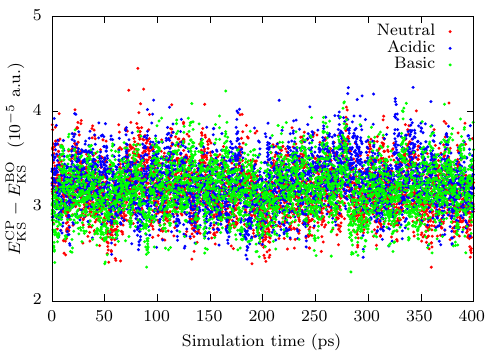}
        \caption{Difference between the Kohn--Sham energy obtained during the 2genCP simulations with $\gamma_L=\gamma_D=0$
	and the BO reference for the three studied systems.}
\label{edif}
\end{figure}

\begin{figure}
	\includegraphics[width=0.5\textwidth]{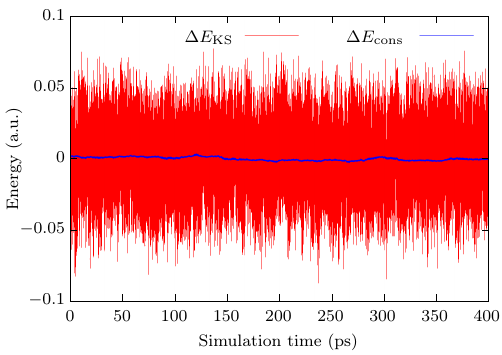}
	\caption{Fluctuations of the Kohn--Sham energy and the conserved quantity (total energy)
	with respect to their average values during the 2genCP simulation with $\gamma_L=\gamma_D=0$
	in the neutral water system.}
\label{econs}
\end{figure}

Thus, apparently the 2genCP method produces in this case a propagation regime that is 
identical to the BO one without even the need to set a non-zero $\gamma_D$.
The following analyses of the forces and energies during the simulations show that it is indeed the case.
Sample structures of the three studied systems
were taken every 0.1~ps
from 
the simulations with $\gamma_L = 0$,
and the atomic forces and Kohn-Sham energy of the system were recalculated 
by completely quenching the wavefunction to the Born--Oppenheimer surface.
As shown in Fig.~\ref{histo},
the distribution of the force modules in these structures 
is practically identical in both the 2genCP and BO cases,
while
the distribution of the module of the force deviation confirms that the differences 
between the CP and BO forces are extremely small, 
practically always well below $5\times 10^{-4}$~a.u.
It is interesting to realize that the extent of these force deviations is much smaller 
than the deviations associated to changing the basis set 
from the most exact one available in CP2K to the infinite limit within the same level of theory,
as seen by comparing Fig.~\ref{histo} here to Fig.~S1 in the SI.
On the other hand, the deviation of the Kohn-Sham energy during the 
2genCP
simulations w.r.t. the BO reference
remains always in the order of $\sim 10^{-5}$ Hartrees (Fig.~\ref{edif}),
with a remarkably good energy conservation (Fig.~\ref{econs}).

It is then clearly demonstrated that the
dynamics provided by the 
2genCP
propagation scheme at the employed level of theory is 
indistinguishable from the Born--Oppenheimer regime even with $\gamma_D = 0$,
as the differences between the 2genCP and BO atomic forces are negligible in practice. 
From this we can conclude that the previous runs with $\gamma_L = 0$ are actually NVE simulations
(note that Eq.~\ref{eq2genCP} reduces to Newton's 2$^\mathrm{nd}$ law when $\gamma_L = \gamma_D = 0$), 
while those with 
$\gamma_L > 0$ are NVT 
runs with different thermostat strengths
as it has been confirmed by analyzing the behavior of the temperature (Sec.~II in the SI).

This remarkable behavior of the 2genCP dynamics 
deserves closing this section with some comments on its relationship
to the employed settings.
The crucial detail that determines 
the ability of the scheme for keeping the system so close to the BO surface
is the number of corrector steps applied, as stated in the original paper.\cite{kuhne_prl2007}
While therein it is not explicitly investigated whether the dissipative behavior depends on the number
of corrector steps, it seems obvious that it must vanish at some point as the system is kept 
closer and closer to the BO surface, which in this case occurs with only 2 corrector steps.
I note that here the choice of 2 corrector steps was not due to intentionally looking for that effect,
but because of its superior accuracy/efficiency balance compared to using
1 corrector step: adding the second step increases the computational cost only a factor of 1.3,
but in conjunction with the ASPC extrapolation 
guarantees an excellent stability of the dynamics. 
Ensuring such stability was necessary in this work, given the long simulation times pursued,
the need for producing a very accurate dynamics in order to properly compute time-dependent properties,
and to guarantee that the fast reorganization of the electronic degrees of freedom
during the H$^+$/OH$^-$ diffusion would remain controlled.

\subsection{Diffusion coefficients via 2$^\mathrm{nd}$-generation CP MD with different thermostats}

After observing that in the studied systems the dynamics provided by the 2genCP method is 
not dissipative, it is straightforward to realize that it is then possible to
use it in conjunction with different thermostats,
since now the necessity of employing a Langevin equation for the propagation scheme has vanished.
Hence, for each of the studied systems I set up three more simulations keeping the previous
settings for the 2genCP propagation 
except the original Langevin thermostat, which is substituted by either
a Nos\'e--Hoover chains thermostat (with the same settings used for the BO-NVT runs),
or a CSVR thermostat (with two different values of the  time constant,
$\tau=100$~fs and $\tau=1000$~fs). 
While Nos\'e--Hoover chains allow a correct estimation of $D$ as seen above,
CSVR~\cite{bussi_jcp2007} can be considered the global version of Langevin (which is local) 
and therefore it affects the dynamics to a much smaller extent, 
allowing a better estimation of time-dependent properties.\cite{bussi_cpc2008}
All these runs were equilibrated for 10~ps, followed by 400~ps of production.
Together with the previous 2genCP simulations within the original Langevin dynamics scheme, 
this makes a total of 7 simulations using 2genCP for each system, 
the results of which I analyze in the following.

The obtained $D$ values are summarized in Table~\ref{difftab} 
and visualized in Fig.~\ref{diffplot}. 
Overall, 
the agreement with the diffusion coefficients estimated via BOMD is excellent, 
with small differences associated to the different thermostats.
The  Nos\'e--Hoover chains thermostat 
produces the same results for the BO and 2genCP regimes (within the statistical errors),
which are also almost equal to those provided by CSVR with $\tau=100$~fs.
The canonical sampling is very well established in these cases, 
judging by the $T$ fluctuations (Figs.~S2--S4 in the SI).
This is also the case of CSVR with $\tau=1000$~fs, 
which produces the correct temperature fluctuations 
even though the average temperature slightly deviates from 300~K,
and its $D$ values perfectly agree with the previous ones when
considering its marked dependence on $T$.
On the other hand, 
Langevin dynamics with $\gamma_L = 10^{-5}$ or $10^{-4}$~fs$^{-1}$ 
has more marked problems to produce the required average temperature, 
being off by about 5~K in at least one of the systems,
and with the temperature fluctuations being below the expected value 
always for $\gamma_L = 10^{-5}$~fs$^{-1}$ and occasionally for $\gamma_L = 10^{-4}$~fs$^{-1}$.
In contrast, Langevin dynamics with $\gamma_L = 10^{-3}$ produces both the correct $\langle T \rangle$
and $\sigma^2_T$ in all systems.
It is interesting to realize that the $D$ values in the $\gamma_L = 10^{-5}$~fs$^{-1}$ case
do nicely agree with those obtained via Nos\'e--Hoover and CSVR thermostats (in addition to all the BO simulations)
when considering the $D(T)$ dependence, while in the $\gamma_L = 10^{-3}$ and $10^{-4}$~fs$^{-1}$ cases
the $D_\mathrm{H_3O^+}$ slightly overestimates that of the other approaches.
It is then quite possible that the latter cases present the well-known problem of the underlying 
stochastic dynamics affecting the estimation of time-dependent properties,
in this case only observed for the (local) Langevin thermostat.

Finally, the simulations with  $\gamma_L = 0$, which are equivalent to the BO-NVE ones,
provide estimations of $D$ which are perfectly in line with those, once more 
reflecting the correlation between the observed $D$ and $\langle T \rangle$.

\subsection{Analyzing the accuracy of $D$ depending on the simulation length}

While it is obviously desirable to achieve the longest simulation times possible,
reaching scales of hundreds of picoseconds at the \textit{ab initio} level as I have done here is a daunting task.
As shown above, methods such as 2genCP may alleviate this cost, 
but even with this approach performing these simulations requires quite some computational effort.
The situation would be much worse in the case that one would request an even higher level of theory,
such as going beyond the pure GGA approximation to include a fraction of HF exchange 
or considering nuclear quantum effects.
\begin{figure}[b!]
\includegraphics[width=0.5\textwidth]{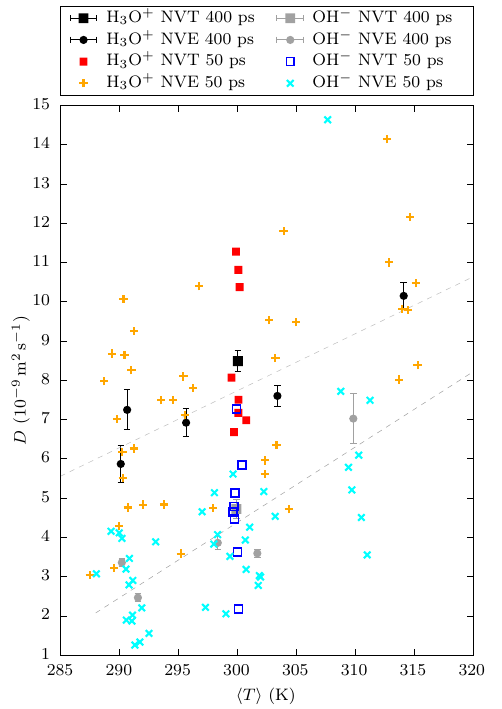}
        \caption{Diffusion coefficients of the solvated excess proton and hydroxide computed using intervals
        of 50~ps of the BOMD runs, in comparison to those obtained with the full 400~ps simulations.}
\label{diffq}
\end{figure}
That would surely limit the affordable simulation time to less than 100~ps, 
probably at most 50 ps, which obviously would very negatively impact on the accuracy of the 
results.
In order to understand such impact, 
I computed the  diffusion coefficients of H$_3$O$^+$(aq) and OH$^-$(aq)
over different parts of the BOMD simulations with lengths of 50~ps (Fig.~\ref{diffq}) 
and 100~ps (Fig.~\ref{diffs}).

As clearly seen in Fig.~\ref{diffq}, 50~ps is an extremely short simulation length to reliably compute the diffusion coefficient
of the solvated proton or hydroxide, with the possible results being dispersed over a huge range.
Even looking only at the results within the NVT ensemble, which show a smaller dispersion than the NVE ones,
it is possible to obtain estimations that are not only quantitatively but also qualitatively wrong 
(i.e. obtaining $D_\mathrm{H_3O^+} < D_\mathrm{OH^-}$).
Increasing the simulation length to 100~ps slightly improves the situation (Fig.~\ref{diffs}), 
and the estimation of $D_\mathrm{H_3O^+}$ via NVT simulations would be now better converged,
while that of $D_\mathrm{OH^-}$ could still be rather off. 
Again, the NVE results show a dispersion that makes impossible to accurately estimate the diffusion coefficients
from single NVE simulations, 
especially if the great dependence of $D$ on $\langle T \rangle$ is ignored 
(I note in passing that the usual procedure of starting a NVE simulation from a state equilibrated via NVT
does not at all guarantee that the average simulation temperature during the NVE run will be the same 
as during the previous equilibration run).
This clearly shows the necessity of averaging over at least several tens of NVE simulations in order to obtain reliable results.

\begin{figure} 
\includegraphics[width=0.5\textwidth]{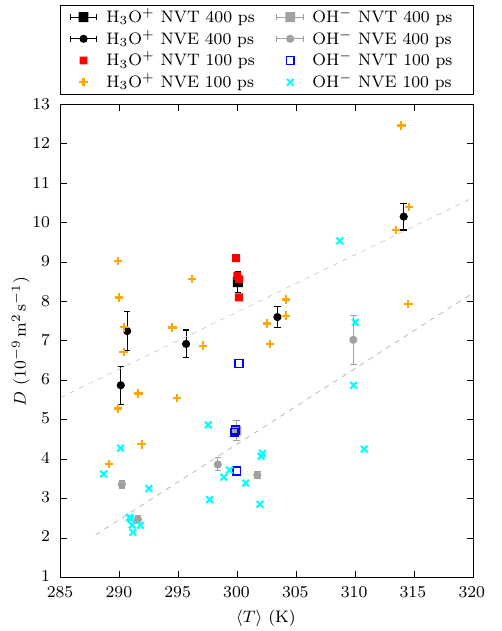}
        \caption{Diffusion coefficients of the solvated excess proton and hydroxide computed using intervals 
        of 100~ps of the BOMD runs, in comparison to those obtained with the full 400~ps simulations.}
\label{diffs}
\end{figure}

\section{Conclusions}

The results above exposed demonstrate that the combination of the RPBE functional with D3 dispersion corrections
provides a very accurate estimation of the diffusion coefficients
of the excess proton and hydroxide ions in water, yielding a
$D_\mathrm{H_3O^+} / D_\mathrm{OH^-}$ ratio that is equal to the experimental one;
in addition to an excellent description of different properties of neutral water such
as its diffusion coefficient, its shear viscosity, and its liquid structure,
in almost perfect agreement with the experimental values.
On the other hand, based on the existing results 
for the similar revPBE-D3 functional,\cite{marsalek_jpcl2017}
it can be expected that improving these RPBE-D3 results
would require adding \textit{both} NQEs and a fraction of HF exchange,
as including only NQEs would produce a (spurious) reduction of $D_\mathrm{H_2O}$  
and probably also of $D_\mathrm{H_3O^+}$ and $D_\mathrm{OH^-}$ 
(since their diffusion mechanisms would be now limited by the
reduced fluctuations of the H--bonding network in such less mobile water),
pushing all three farther from the experimental values.
On the other hand, 
it is difficult to predict how the interplay of HF exchange and NQEs on RPBE-D3 would affect
the absolute vales of $D_\mathrm{H_3O^+}$ and $D_\mathrm{OH^-}$.
In any case --and even if partly due to such error compensation--,
the fact that RPBE-D3 performs so well for all the three species H$_2$O, H$_3$O$^+$ and OH$^-$ 
at an affordable computational cost compared to much more expensive approaches,
and without resorting to other approximations usually taken in the field 
(e.g. increasing the simulation temperature),
makes it the best choice so far for simulating them.

Regarding the simulation strategy, the 2$^\mathrm{nd}$-generation Car--Parrinello MD 
method yields excellent results
that are indistinguishable from those obtained with regular Born--Oppenheimer MD
at a smaller computational cost.
Actually, 2genCP provides in this case a propagation that is so close to the Born--Oppenheimer regime
that it is not necessary to employ the Langevin dynamics scheme in which it was originally formulated,
being possible to use instead other regular but global thermostats such as Nos\'e--Hoover chains or CSVR,
both of which improve the accuracy of the computed time-dependent properties over Langevin dynamics.

The statistical accuracy of the results presented here is strongly backed-up by the fact that the many simulations performed,
employing different propagation regimes and thermostat settings 
(which guarantees that the different simulations are completely decoupled),
provide essentially the same results when taking into account the statistical uncertainties
and especially the strong dependence of the diffusion coefficients on the simulation temperature,
with only minor differences associated to the different thermostat schemes employed.
Moreover, I show the huge importance of reaching simulation times in the order of hundreds of picoseconds
when computing the diffusion coefficients of the solvated proton and hydroxide.
At that respect, it must be recognized that 
the times of 400~ps per simulation achieved in this work are much longer
than those seen in most of the \textit{ab initio} MD studies in the literature,
where it is usual to find estimations of the diffusion coefficients of 
these species
with simulation times in the order of 50--100~ps. 
However, as I have just shown, these would be indeed too short 
to obtain reliable values of 
$D_\mathrm{H_3O^+}$ and $D_\mathrm{OH^-}$,
being rather possible to obtain qualitatively wrong results.
Moreover, special care must be taken when computing diffusion coefficients from NVE simulations,
since it is necessary to average over at least tens of them in order to obtain reliable estimations,
and in particular the great dependence of $D$ on $T$ must be acknowledged.

Finally, I close this paper with some 
outlook regarding the
underlying structural diffusion mechanism of 
the H$_3$O$^+$(aq) and OH$^-$(aq) ions. An initial analysis of the obtained trajectories
supports the currently standing picture of both the preferred solvation structures
of both ions and their migration mechanisms, 
i.e. the `traditional Grotthuss mechanism'~\cite{marx_cpc2006}
in the case of H$_3$O$^+$(aq) and the `dynamical hypercoordination mechanism'~\cite{marx_cpc2006,marx_chemrev2010}
for OH$^-$(aq).
However, deeper analyses of the vast amount of generated data are underway. 
Such analyses, which are out of the scope of this paper and will be unfolded in following publications, 
are aimed at understanding more involved aspects of the diffusion mechanism of
the excess proton and hydroxide,
such as the precise determination of energetic barriers for proton transfer, 
the occurrence and timescale of concerted proton transfer events,
and the landscape of the different solvation structures of both species.

\section*{Supporting Information}

The SI document includes a detailed account
of the tests for validating the employed basis set,
as well as detailed plots showing the evolution of the temperature
and its fluctuations during the different simulations,
and the MSD computed in all the simulations.

\section*{Acknowledgments}

It is my pleasure to thank Jorge Kohanoff, Emilio Artacho, Dominik Marx and Thomas K\"uhne for insightful discussions,
as well as Alex Hannon and Alan Soper for kindly providing the experimental data.
I acknowledge PRACE for awarding me access to the MareNostrum4 supercomputer at 
the Barcelona Supercomputing Center (Spain).
This project has received funding from the European Union's Horizon 2020 research and innovation
programme under the Marie Sk\l{}odowska-Curie grant agreement No 748673.
This work has been supported by the Madrid Government (Comunidad de Madrid-Spain) under the Multiannual Agreement 
with Universidad Polit\'ecnica de Madrid 
in the line Support for R\&D projects for Beatriz Galindo researchers, 
in the context of the V PRICIT (Regional Programme of Research and Technological Innovation).


\bibliography{munoz_R}

\begin{thebibliography}{85}
\bibitem [{not({\natexlab{b}})}]{note_erg}%
  \BibitemOpen
  \href@noop {} {\  ({\natexlab{b}})},\ \bibinfo {note} {{A} strict proof of
  this assumption would require dwelling on fundamental aspects such as the
  ergodicity of NVE simulations of these finite-size aqueous systems, which is
  far out of the scope of this paper. I note that, were this assumption false
  (i.e. these simulations being non-ergodic in the strict sense), that would
  imply that reliably determining $D_\mathrm{H_3O^+}$ or $D_\mathrm{OH^-}$ via
  single NVE simulations is impossible.}\BibitemShut {Stop}%
\end{thebibliography}

\clearpage



\section*{Supplementary material}

# Accurate diffusion coefficients of the excess proton and hydroxide in water via extensive *ab initio* simulations with different schemes

Daniel Muñoz-Santiburcio,[1,2]

[1] *CIC nanoGUNE BRTA, Tolosa Hiribidea 76, 20018 San Sebastián, Spain*
[2] *Instituto de Fusión Nuclear "Guillermo Velarde", Universidad Politécnica de Madrid,
C/ José Gutiérrez Abascal 2, 28006 Madrid, Spain*

## I. VALIDATION OF THE BASIS SET

In order to check the validity of the basis set, I employed two test systems consisting of 32 water molecules in a cubic unit cell with $a = 9.87$ Å in which a proton was added or removed to obtain respectively an acidic and basic bulk water system. In order to generate an ensemble of representative structures for these systems, two respective AIMD runs were carried out in the NVT ensemble within the 2genCP framework, employing a 0.4 fs timestep and a Nosé–Hoover chains thermostat at 300 K (thus equal to the 2genCP runs with Nosé–Hoover chains described in the main text). The electronic structure of the system was treated as described in the main text for all the systems (i.e. using GTH pseudopotentials together with `TZV2P` basis sets, a 500 Ry cutoff and using the RPBE-D3 functional). The systems were equilibrated during 10 ps, followed by production runs of 100 ps. Sample structures were taken every 1.0 ps, obtaining 101 structures for each system.

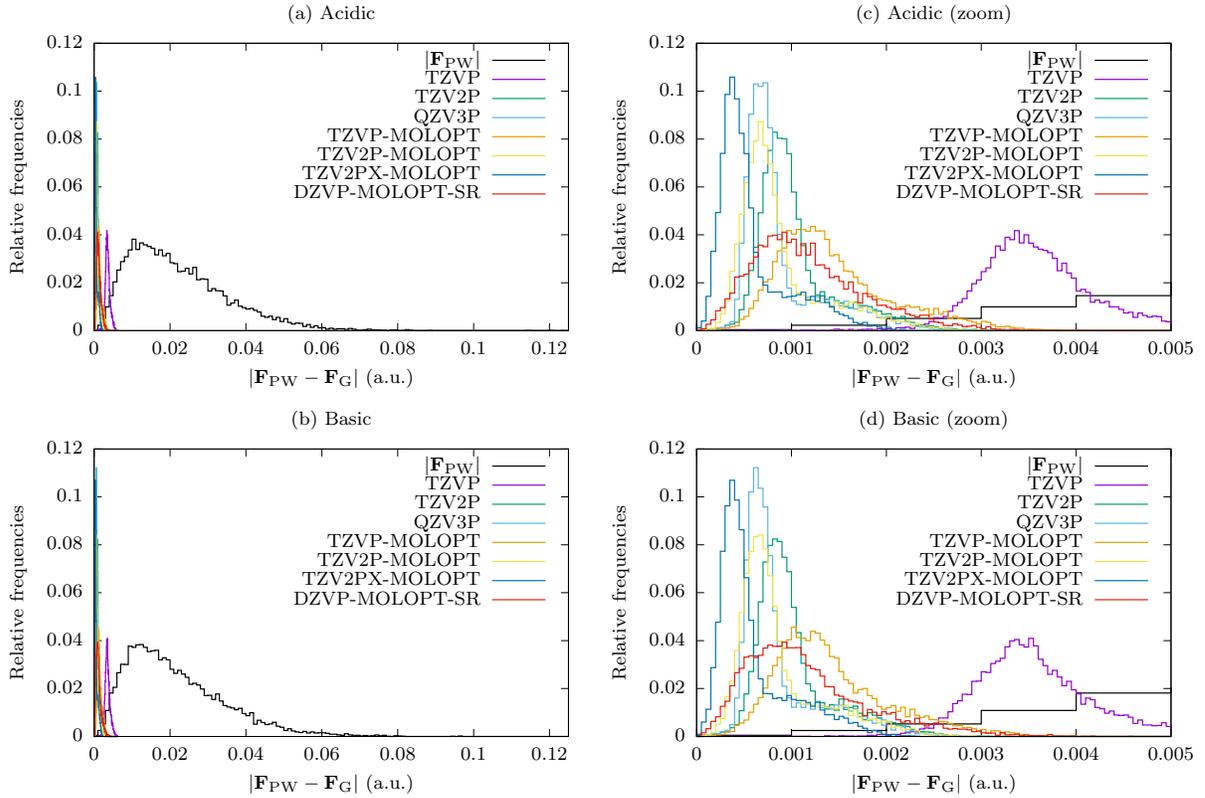

Figure S1. Distributions of the absolute atomic forces computed at the planewave level with an almost infinite basis set, $|\mathbf{F}_{PW}|$, computed over a set of 101 snapshots of a 32 water system plus or minus a proton (resp. (a) and (b)), in comparison to the distribution of the module of the deviations of the atomic forces with different Gaussian basis sets with respect to the 'exact' forces at the planewave limit, $|\mathbf{F}_{PW} - \mathbf{F}_G|$, being G one of `TZVP`, `TZV2P`, etc. (c) and (d) show a magnification of the region where the force deviations lie.

Table S1. Statistical descriptors of the distributions of the module of the force deviations with the different Gaussian basis sets with respect to the infinite planewave reference, (i.e. $|\mathbf{F}_{\text{PW}} - \mathbf{F}_{\text{G}}|$ where G is one of `TZVP`, `TZV2P`, etc.), ordered by the mean values of each distribution.

| Acidic water | | | |
|---|---|---|---|
| Mean | Median | 3$^{\text{rd}}$ quartile | Basis set |
| 0.0005953 | 0.0004636 | 0.0007264 | `TZV2PX-MOLOPT` |
| 0.0008906 | 0.0007442 | 0.0010235 | `TZV2P-MOLOPT` |
| 0.0008952 | 0.0007397 | 0.0009787 | `QZV3P` |
| 0.0010506 | 0.0009316 | 0.0011672 | `TZV2P` |
| 0.0011749 | 0.0010714 | 0.0015067 | `DZVP-MOLOPT-SR` |
| 0.0014494 | 0.0013161 | 0.0017476 | `TZVP-MOLOPT` |
| 0.0035878 | 0.0035174 | 0.0039176 | `TZVP` |
| Basic water | | | |
| Mean | Median | 3$^{\text{rd}}$ quartile | Basis set |
| 0.0006033 | 0.0004645 | 0.0007486 | `TZV2PX-MOLOPT` |
| 0.0008711 | 0.0007285 | 0.0010122 | `TZV2P-MOLOPT` |
| 0.0008751 | 0.0007189 | 0.0009752 | `QZV3P` |
| 0.0010264 | 0.0009119 | 0.0011467 | `TZV2P` |
| 0.0011553 | 0.0010232 | 0.0014658 | `DZVP-MOLOPT-SR` |
| 0.0014208 | 0.0012955 | 0.0017301 | `TZVP-MOLOPT` |
| 0.0035655 | 0.0034918 | 0.0039091 | `TZVP` |

With these sets of structures, the atomic forces were recomputed with CP2K by completely quenching the wavefunction to the Born–Oppenheimer surface using the following basis sets: `TZVP`, `TZV2P`, `QZV3P`, `TZVP-MOLOPT`, `TZV2P-MOLOPT`, `TZV2PX-MOLOPT`, and `DZVP-MOLOPT-SR`. For this, I kept intact the rest of the electronic structure settings except the D3 dispersion correction, which was removed in order to facilitate the comparison between the different basis sets. In addition, in order to obtain a valid reference for the 'exact' forces within the employed level of theory, I used the planewave code Abinit. This code allows to use the very same GTH pseudopotentials as CP2K, which together with a PW basis set with a cutoff of 900 Rydberg provides in practice the limit of an infinite basis set.

After computing all the individual forces, the appropriateness of the Gaussian basis sets was checked by computing the module of the deviation of the forces with the Gaussian basis w.r.t. the forces computed with the planewave basis, i.e. $|\mathbf{F}_{\text{PW}} - \mathbf{F}_{\text{G}}|$. The corresponding distributions are shown in Fig. S1, where in addition to the histograms of the module of the deviations it is also shown the histogram of the absolute force modules computed at the PW level $|\mathbf{F}_{\text{PW}}|$. Some statistical descriptors of all the distributions are listed in Table S1 to facilitate the ranking of the different basis sets in order of accuracy.

It is clearly seen that the force deviations for most of the Gaussian basis sets are very small compared to the absolute exact forces, being `TZVP` the only exception which performs clearly much worse than the rest. From the most to the less accurate set, they can be ranked as:

`TZV2PX-MOLOPT` > `TZV2P-MOLOPT` ∼ `QZV3P` > `TZV2P` > `DZVP-MOLOPT-SR` > `TZVP-MOLOPT` ≫ `TZVP`.

On the other hand, the computational cost for each basis would rank them as

`TZV2PX-MOLOPT` > `TZV2P-MOLOPT` > `TZVP-MOLOPT` ≫ `QZV3P` > `TZV2P` > `DZVP-MOLOPT-SR` > `TZVP`.

In view of these results, the chosen basis for the full study of the diffusion coefficients of $H_2O$, $H_3O^+$ and $OH^-$ was `TZV2P` which showed a very good balance between accuracy and computational cost.

## II. EVOLUTION OF THE TEMPERATURE DURING THE SIMULATIONS

The dependence of the diffusion coefficients on the temperature of the system, in addition to the necessity of checking whether a proper sampling is being established in the 2genCP simulations, make it necessary to closely monitor the behavior of the temperature in all the cases. In Figs. S2, S3 and S4 I show the running average of the temperature $\langle T \rangle$ and the evolution of its variance $\sigma_T^2$ during the simulations for the neutral, acidic and basic water systems, respectively. The extent in which a proper canonical sampling is being achieved in a given NVT simulation can be checked by comparing the observed $\sigma_T^2$ to its theoretical value in a NVT ensemble, i.e. $\sigma_T^2(\text{NVT}) = \frac{2}{3N}\langle T \rangle^2$ [1]. This value for $T = 300$ K is represented in Figs. S2, S3 and S4 with a dashed black line.

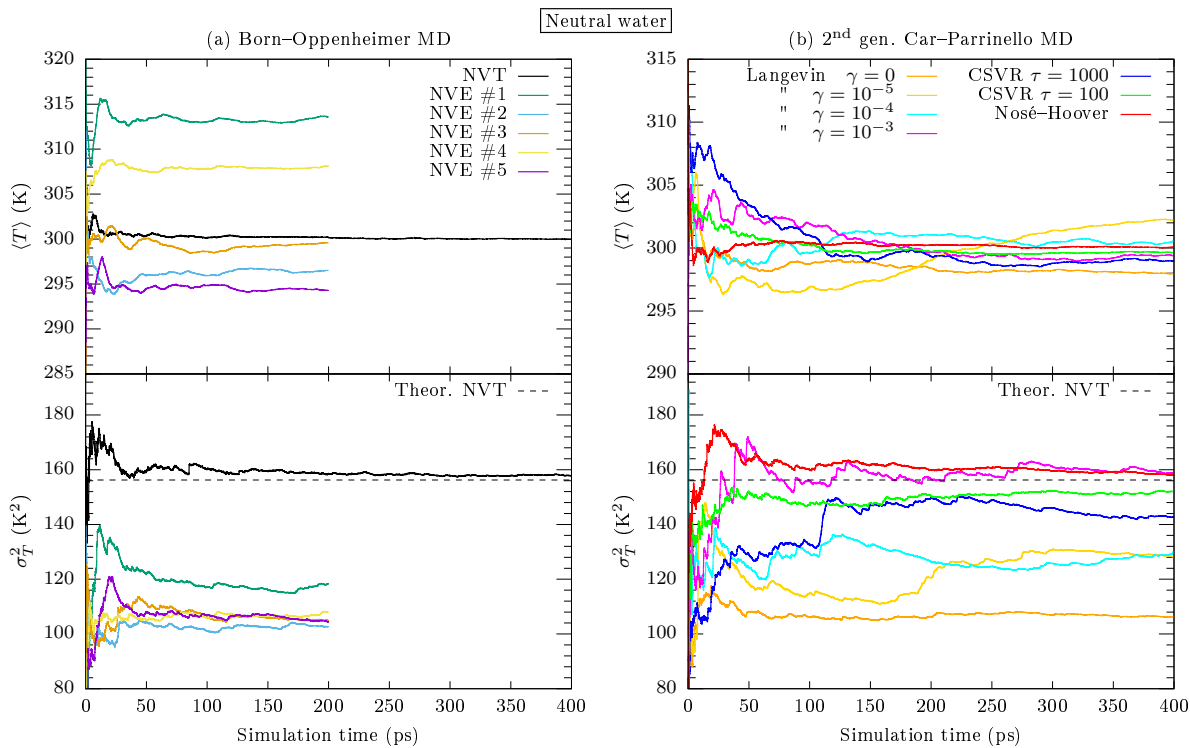

Figure S2. Evolution of the temperature in the simulations of the neutral water system.

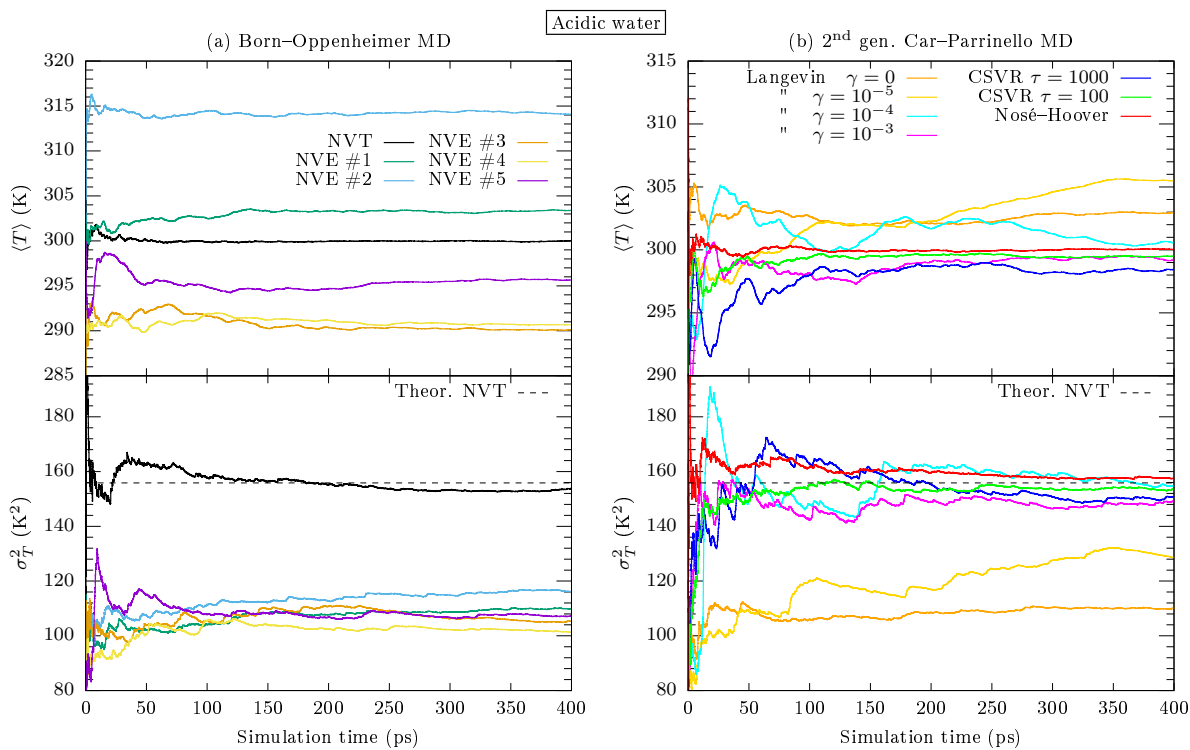

Figure S3. Evolution of the temperature in the simulations of the acidic water system.

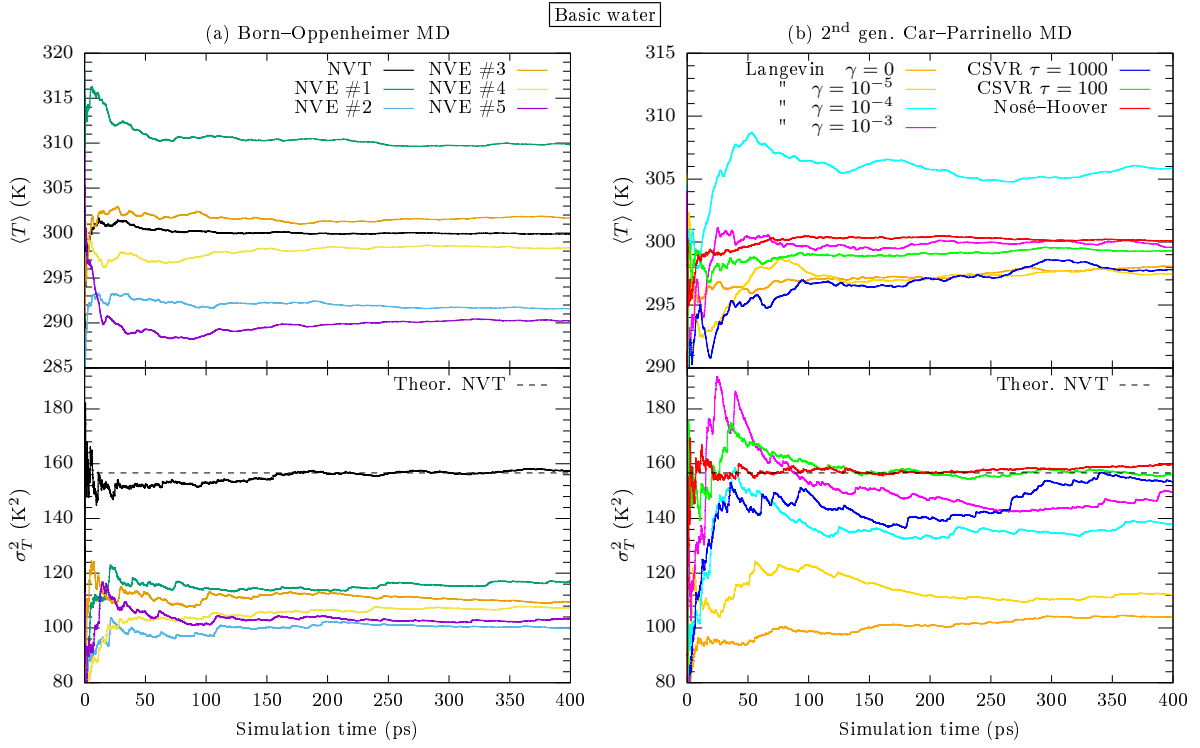

Figure S4. Evolution of the temperature in the simulations of the basic water system.

## III. MSD COMPUTED IN THE DIFFERENT SIMULATIONS

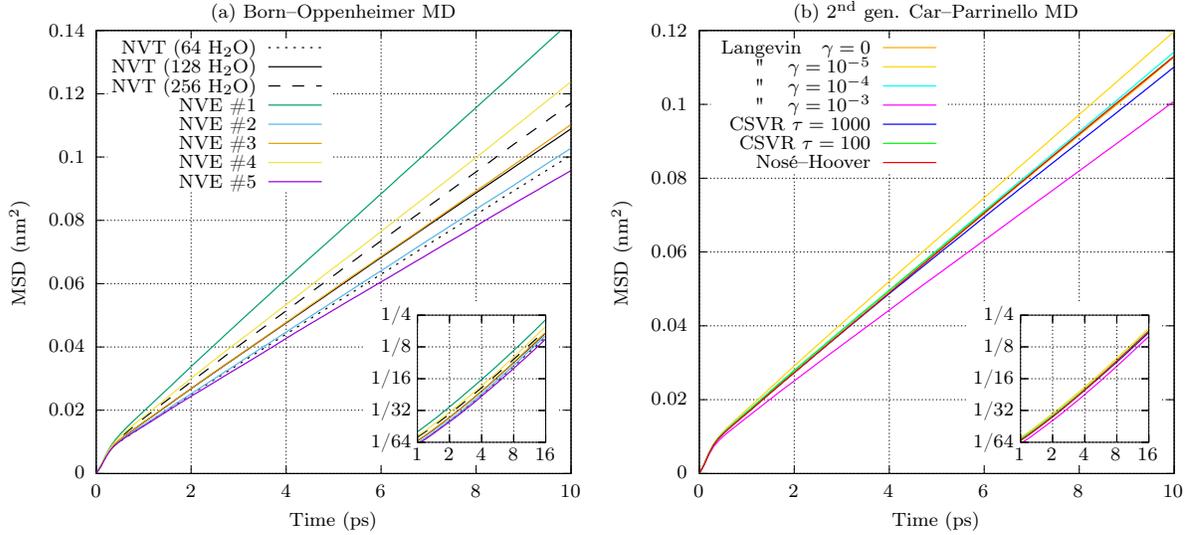

Figure S5. MSD($t$) obtained in the simulations of the neutral water system, including the 64 and 256 $H_2O$ systems used for determining the finite-size effects. The insets show the same MSD($t$) with logarithmic scales in both axes, within a different range but with the same units as the main plots. The MSD is computed using only the position of the O atom in each $H_2O$ molecule, and averaging over all the molecules in the system. The diffusion coefficients $D_{H_2O}$ are computed using the slope of the MSD in the range $2 \leq t \leq 10$ ps.

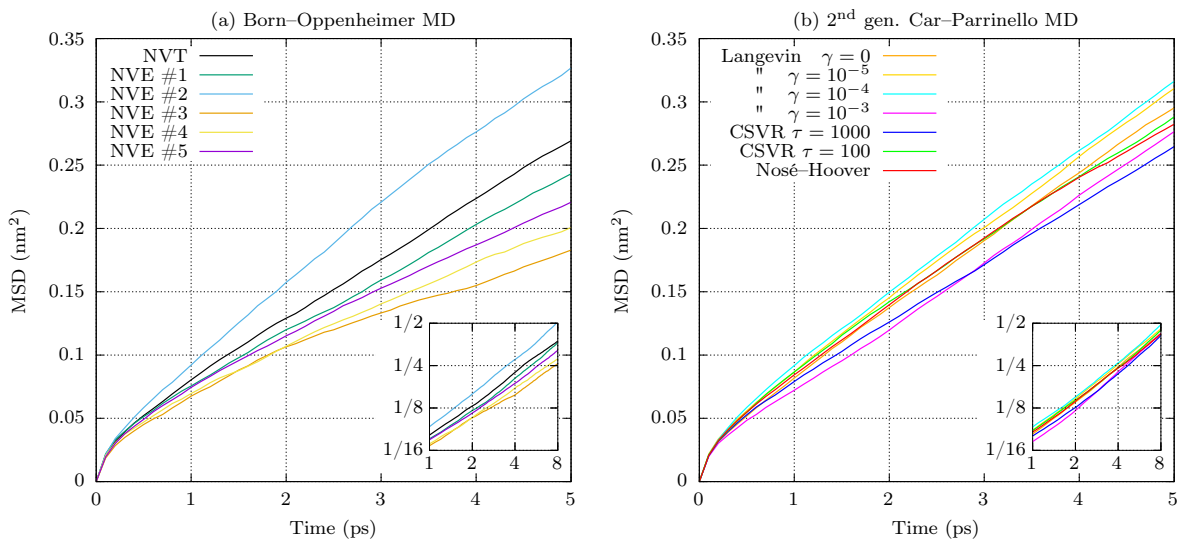

Figure S6. MSD($t$) obtained in the simulations of the acidic water system, computed using only the position of the O atom in each $H_3O^+$ ion. The insets show the same MSD($t$) with logarithmic scales in both axes, within a different range but with the same units as the main plots. The diffusion coefficients $D_{H_3O^+}$ are computed using the slope of the MSD in the range $2 \leq t \leq 5$ ps.

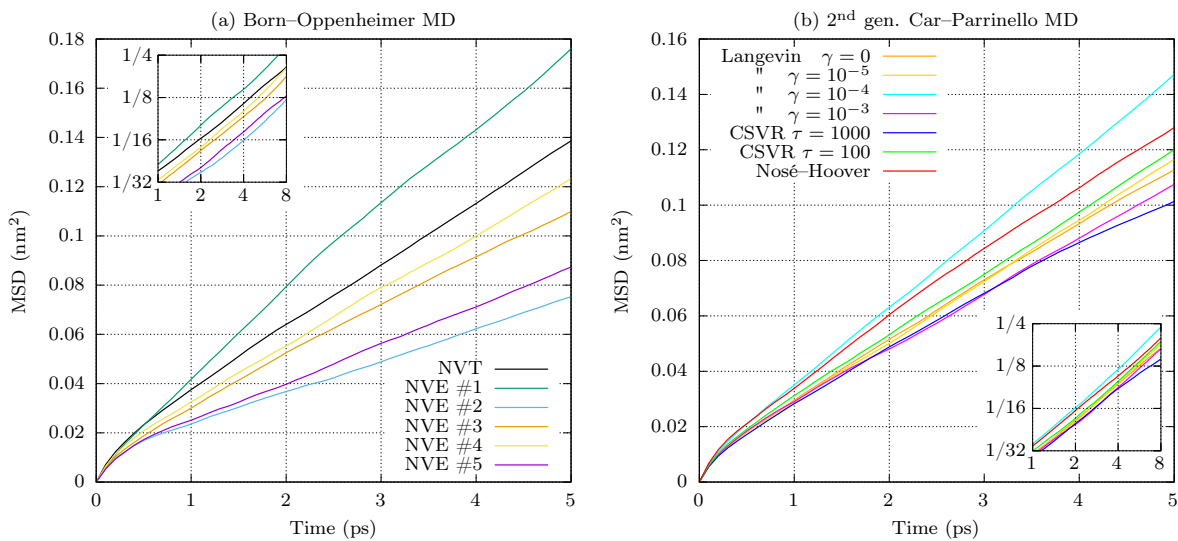

Figure S7. MSD($t$) obtained in the simulations of the basic water system, computed using only the position of the O atom in each $OH^-$ ion. The insets show the same MSD($t$) with logarithmic scales in both axes, within a different range but with the same units as the main plots. The diffusion coefficients $D_{OH^-}$ are computed using the slope of the MSD in the range $2 \leq t \leq 5$ ps.



\end{document}